\documentclass[conference]{IEEEtran}
\IEEEoverridecommandlockouts

\usepackage{cite}
\usepackage{amsmath,amssymb,amsfonts}
\usepackage{algorithmic}
\usepackage{graphicx}
\usepackage{textcomp}
\usepackage{xcolor}
\usepackage{enumitem}
\usepackage{multirow}
\usepackage{booktabs}
\usepackage[hidelinks]{hyperref}
\usepackage{breakurl}
\newcommand{\makecell}[1]{\begin{tabular}{@{}c@{}}#1\end{tabular}}

\makeatletter
\def\url@deffont{\normalfont}
\makeatother

\title{Robustness and Trade-offs for Code LLMs on Protected Code}
\author{
\IEEEauthorblockN{Jin Wen\IEEEauthorrefmark{1},
Yuejun Guo\IEEEauthorrefmark{2},
Yujie Ma\IEEEauthorrefmark{3}\IEEEauthorrefmark{4}\thanks{\IEEEauthorrefmark{4} Corresponding author.},
Qiang Hu\IEEEauthorrefmark{3},
Maxime Cordy\IEEEauthorrefmark{1}}
\IEEEauthorblockA{\IEEEauthorrefmark{1}University of Luxembourg, Luxembourg, \{jin.wen,maxime.cordy\}@uni.lu}
\IEEEauthorblockA{\IEEEauthorrefmark{2}Luxembourg Institute of Science and Technology, Luxembourg, yuejun.guo@list.lu}
\IEEEauthorblockA{\IEEEauthorrefmark{3}Tianjin University, China, yujiema0911@tju.edu.cn, qianghu0515@gmail.com}
}

\begin{document}
\maketitle

\begin{abstract}
Code large language models (LLMs) are increasingly used on software artifacts that may be intentionally obfuscated for intellectual-property protection, reverse-engineering resistance, or controlled access.
In reverse engineering and security analysis, deobfuscation is commonly treated as the preprocessing step before downstream analysis or inference~\cite{udupa2005deobfuscation}, yet its utility for code LLM pipelines has not been systematically validated across models and protection methods.
We present an execution-based study of seven code LLMs on protected-code translation and completion, spanning source programs from C++, Go, Java, and JavaScript, five obfuscation methods, and three inference protocols: plain, obfuscated, and deobfuscated. Our results show that direct inference on obfuscated code often matches or exceeds inference on restored code. In this controlled benchmark, higher-capability models such as GPT-4.1 and Qwen3-Coder-30B retain about 90\% Pass@1 on obfuscated translation inputs, indicating that explicit restoration is often unnecessary. Same-model restoration does recover some obfuscation-induced failures, but it also degrades many cases that already succeed, with lower-capability models showing the largest net losses. Across settings, model capability is the primary factor, while source language and obfuscation method have secondary but consistent effects. Overall, our findings support model-aware pipeline design and indicate that protected-code workflows should be evaluated primarily with execution-based metrics rather than static similarity alone.
\end{abstract}


\section{Introduction}
Despite the rapid evolution of software systems, many organizations remain dependent on legacy code that is critical to business operations.
Such systems are often difficult to maintain due to missing documentation, outdated technology stacks, and gaps in institutional knowledge.
At the same time, software vendors and developers commonly employ code obfuscation to protect intellectual property (IP) and prevent unauthorized code analysis or intellectual theft.
% Obfuscation is widely adopted in industry to conceal proprietary logic and raise the cost of reverse engineering, and it is a standard component of software protection suites such as ProGuard.
Obfuscation is widely adopted in industry to conceal proprietary logic and raise the cost of reverse engineering, and is commonly integrated into software protection suites such as ProGuard~\cite{proguard_manual}.

Traditional code obfuscation transforms a program into a semantically equivalent but significantly harder-to-understand form.
Common transformation techniques range from identifier renaming and control flow restructuring to dead code insertion~\cite{collberg1997taxonomy,schrittwieser2016softwareobfuscation}, all designed to increase the cognitive and computational burden on both human analysts and automated tools.
The foundational assumption of classical obfuscation research is that the transformed code remains executable while being substantially more resistant to interpretation or modification.
In this paper, we use \emph{protected code} as the broader pipeline-level term for code transformed to restrict inspection or reuse, and \emph{obfuscated code} for the concrete transformed artifacts produced by the protection methods we evaluate. Since all studied protection methods operate through obfuscating transformations, we use the two terms interchangeably when referring to model inputs.

The operational setting we target is authorized maintenance, migration, auditing, or incident response on protected artifacts, such as externally packaged, controlled-access, or legacy code. We do not study adversarial IP circumvention. The reliability question is whether adding restoration to such a workflow improves the probability of correct downstream behavior.

With the widespread use of large language models (LLMs) for software engineering tasks such as code completion, translation, and summarization, a new set of challenges have emerged~\cite{du2024evaluating,pan2024lost}.
Unlike traditional static analysis tools~\cite{moller2012static,venkatesh2024emergence}, LLMs offer deep semantic understanding of code patterns and can infer deep program behaviour directly from surface representations, without requiring explicit structural analysis.
This capability raises a central question: \textit{can LLMs still extract meaningful semantics from obfuscated code and perform downstream software engineering tasks on it?} And more broadly, \textit{are traditional obfuscation techniques still effective against such powerful neural code models, or must they be revisited and strengthened?}

In response to this landscape, recent work such as CodeCipher proposes a new form of LLM-aware code obfuscation~\cite{codecipher}.
Rather than simply making code harder for humans to read or traditional analysis tools to examine, CodeCipher learns token perturbations in the model’s embedding space that aim to hide sensitive code features while preserving the model’s functional outputs on tasks like code completion, translation, and summarization.
While such approaches highlight the growing need for privacy-centric code transformations in the LLM era, the field currently lacks a comprehensive empirical understanding of which obfuscation methods actually hinder LLM-based code understanding and reuse, how different transformation families compare in their effectiveness, and how factors such as model architecture, scale, and programming language influence robustness.

In this paper, we present a systematic empirical study of the interplay between code obfuscation and large language models for code, evaluated through an execution-based framework across seven code LLMs, four source languages, and five protection methods, using code translation as the primary evaluation setting. Our contributions are threefold. First, we provide an execution-based comparison of plain, obfuscated, and deobfuscated inference across the above settings. Second, we reframe deobfuscation as a rescue--degradation trade-off rather than treating it as uniformly beneficial, and operationalize this trade-off through conditional metrics grounded in executable outcomes. Third, we derive operational guidance for protected-code pipelines by identifying where model capability dominates and where language and protection method act as secondary boundary conditions.

\section{Related Work}
We review work on code obfuscation and protection, code LLMs, LLM-based deobfuscation, and robustness under semantics-preserving transformations.

\subsection{Code Obfuscation}
Software obfuscation transforms code into forms that are harder to understand while preserving behavior~\cite{obfuscation}. It is used to protect intellectual property, deter tampering, and raise reverse-engineering cost. Standard transformations include identifier renaming, control-flow changes, and data encoding, implemented by tools such as ProGuard, DashO, and Dotfuscator. Traditional research studies the resulting static and dynamic analysis hurdles and provides taxonomies of transformation families.

\subsection{LLMs for Code Understanding and Generation}
Recent studies evaluate LLMs on completion, translation, summarization, and error detection. Execution-based benchmarks such as HumanEval measure functional correctness rather than surface similarity. These studies show that LLMs can synthesize executable code, but also reveal reliance on names and memorized patterns when surface cues change~\cite{le2025names,zhang2025unseen,djire2026memorization}.

\subsection{LLMs and Code Obfuscation / Deobfuscation}
With the rise of powerful LLMs, researchers have begun exploring their capacity to reverse or mitigate obfuscation.
A number of studies investigate LLM-based deobfuscation in both source code and binary contexts. For example, early work explores whether general LLMs can reduce code complexity under multiple source-to-source obfuscations, showing that models can learn to simplify or "clean" obfuscated code and outperform compiler-based optimizers in certain settings~\cite{beste2025exploring,jiang2026cascade}.
In the binary realm, systematic evaluations such as Deconstructing Obfuscation examine the performance of commercial LLMs on assembly-level deobfuscation, proposing frameworks that categorize obfuscation resistance and expose distinct error patterns across models~\cite{tkachenko2025deconstructing}.
Recent benchmarks like BinDeObfBench evaluate LLMs on pseudocode deobfuscation across diverse transformations and highlight the importance of reasoning capabilities and task-specific training over scale alone~\cite{hu2026can}.
These works demonstrate that LLMs can assist in deobfuscation tasks but also reveal substantial variation in success depending on transformation type and model traits.

In contrast to traditional obfuscation, recent research proposes obfuscation techniques explicitly targeting learned models.
For instance, CodeCipher introduces LLM-aware transformations that perturb token representations in embedding space to conceal sensitive code features while preserving model outputs on tasks like completion and translation.
This approach illustrates a shift toward privacy-centric obfuscation for neural models, and raises questions about how to empirically assess the effectiveness of such transformations across models and tasks.
While promising, existing work on LLM-aware obfuscation does not yet offer a comprehensive empirical comparison across multiple obfuscation families, model types, and programming languages.

\subsection{Robustness and Semantics-Preserving Transformations}
Beyond deobfuscation, several recent empirical studies examine how code obfuscation affects LLM performance on downstream tasks.
A systematic investigation into obfuscation effects on LLM-based vulnerability detection reveals that various obfuscation classes (e.g., layout, data flow, control flow) can both harm and, in some cases, improve model performance depending on how they alter surface cues. These dual effects underscore the complex interplay between obfuscation and neural reasoning, and highlight the need for robust evaluation frameworks~\cite{li2025systematic,zhang2025unseen}.
Another line of work adopts controlled obfuscation as a structured test of semantic understanding, showing that performance declines as obfuscation complexity increases and that general-purpose models can exhibit resilience compared to code-focused counterparts~\cite{nikiema2025code,pan2026readability,yang2026membership}.
These studies provide empirical baselines for semantic comprehension under obfuscation and inform the design of robustness benchmarks~\cite{abdelsalam2026confused}.

\section{Problem Statement and Research Questions}
\label{sec:rqs}

\begin{figure*}[!ht]
    \centering
    \includegraphics[width=\textwidth]{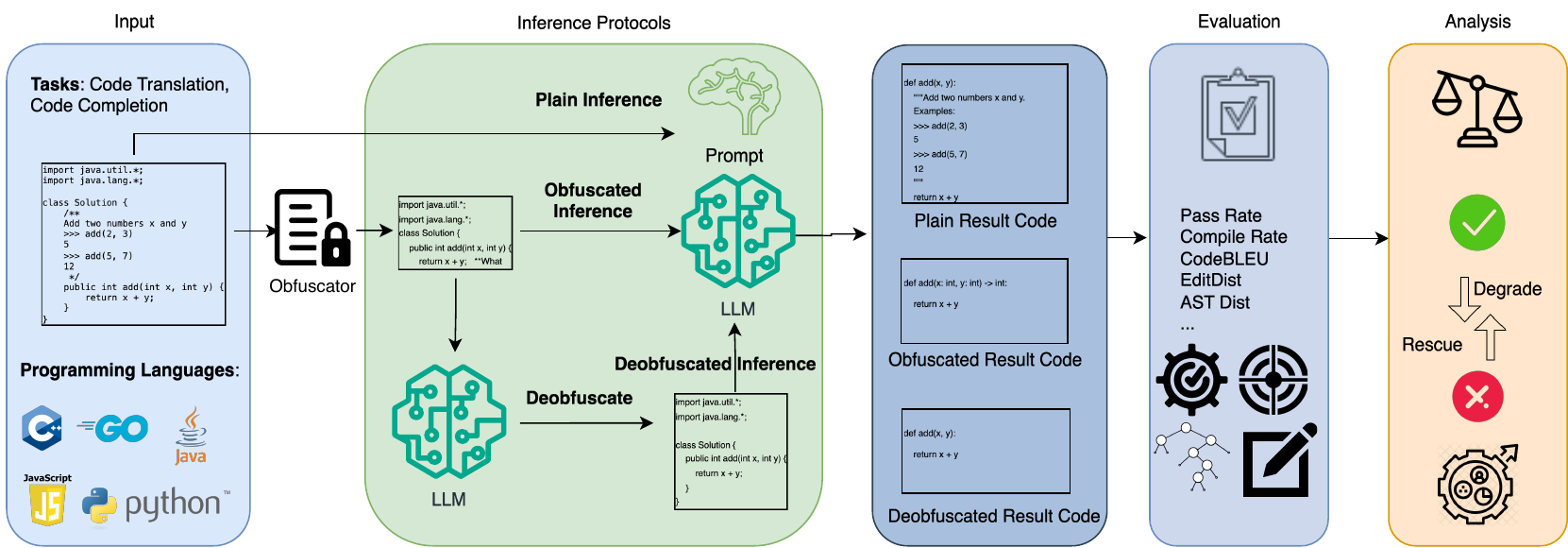}
    \caption{Workflow diagram illustrating the different stages of protected-code processing. Python is the target language for translation and the language of the completion benchmark, not a translation source language.}
    \label{fig:workflow}
\end{figure*}

Despite the long history of code obfuscation for IP protection and reverse-engineering resistance, it remains unclear how such transformations interact with modern LLMs in real software engineering workflows.
Prior work on traditional obfuscation has focused on static and dynamic analysis~\cite{udupa2005deobfuscation}, while studies on LLM deobfuscation suggest that models can sometimes recover semantics from obfuscated code~\cite{hu2026can}.
Work on robustness under semantic-preserving transformations exists as well~\cite{pan2026readability,yang2026membership}, yet these studies generally do not connect transformation effects to execution-based outcomes on practical tasks such as code translation. Figure~\ref{fig:workflow} summarizes the workflow setting studied in this paper.

This gap is consequential. LLMs are increasingly embedded in software engineering pipelines, and practitioners commonly assume that deobfuscation improves model performance on protected code. This assumption, however, has not been systematically evaluated across models, languages, and obfuscation methods. It remains unclear when deobfuscation genuinely helps, when it degrades performance, and how practitioners should make principled pipeline decisions in the presence of such uncertainty.

To address these gaps, we frame our study around four empirical research questions:

\begin{itemize}[leftmargin=*, nosep]
    \item \textbf{RQ1 (Direct vs.\ Restored):} Can code LLMs operate effectively on protected code without restoration, or does a deobfuscation step improve outcomes?
    \item \textbf{RQ2 (Rescue vs.\ Degradation):} When deobfuscation is applied, which cases benefit (rescued) and which are harmed (degraded)?
    \item \textbf{RQ3 (Boundary Conditions):} How do model capability, source language, and protection method jointly shape the rescue–degradation trade-off?
    \item \textbf{RQ4 (Practical Guidance):} What decision rules should guide pipeline selection for production systems handling protected code?
\end{itemize}

\section{Experimental Setup}
\label{sec:experiments}
% put the implementation environment here. For instance, All experiments were conducted on ....

\subsection{Tasks and Data}
We study how source-level obfuscation and subsequent restoration affect code LLM performance on two downstream tasks: \textbf{code translation} and \textbf{code completion}. Translation serves as the primary empirical axis, as it requires the model to process the full source program and produce a functionally equivalent output. Completion is included as supporting evidence to assess whether obfuscation effects generalize across task types.

For both tasks, we use HumanEval-style problems with 164 instances per dataset setting. This gives a controlled function-level benchmark with executable tests, but it does not model repository-scale dependencies, build systems, or long-context maintenance tasks. For code completion, the input is the standard HumanEval prompt. For code translation, we use multilingual HumanEval extensions in which the source program is provided in one of four languages (C++, Go, Java, or JavaScript) and the model must produce a functionally equivalent Python code. In translation, the model receives the full source-language program, including the prompt and reference implementation when available, and is evaluated on whether the translated output is a Python function with the required entry point.

\subsection{Models}
We evaluate seven code LLMs spanning multiple model families and parameter scales:
\begin{itemize}[leftmargin=*, nosep]
    \item CodeLlama-7B, CodeLlama-70B~\cite{roziere2023code}
    \item DeepSeek-Coder-V2~\cite{zhu2024deepseek}, DeepSeek-R1-Qwen-14B~\cite{guo2025deepseek}
    \item Qwen2.5-Coder-14B~\cite{hui2024qwen2}, Qwen3-Coder-30B~\cite{qwen2025qwen3coder30b}
    \item GPT-4.1~\cite{openai2025gpt41}
\end{itemize}

This selection covers open-source and proprietary models, instruction-tuned and reasoning-oriented variants, enabling analysis of how model capability interacts with obfuscation robustness.

\subsection{Obfuscation and Restoration}
We apply five protection methods covering surface- and structure-level transformations:
\begin{itemize}[leftmargin=*, nosep]
    \item Identifier Rename: Identifier renaming replaces meaningful variable and method names with meaningless identifiers to hinder code comprehension while preserving program semantics~\cite{cimato2005overcoming}.
    \item Dead Branch Injection: Dead branch injection introduces syntactically valid but semantically unreachable code paths to increase program complexity without altering runtime behavior~\cite{collberg1997taxonomy}.
    \item Remove Symbols: Removing non-functional elements such as comments and whitespace reduces code readability without affecting program execution~\cite{nagra2009surreptitious}.
    \item Random: Random identifier renaming maps detected identifiers to seed-controlled meaningless names while preserving program structure~\cite{hort2025semantic}.
    \item CodeCipher: CodeCipher performs LLM-aware obfuscation by learning transformations that preserve semantics while reducing the effectiveness of language model-based code understanding~\cite{codecipher}.
\end{itemize}

In the artifact, Identifier Rename maps detected identifiers to deterministic names (\texttt{v0}, \texttt{v1}, \ldots), Random maps identifiers to unique seed-controlled alphanumeric names, Dead Branch Injection inserts unreachable junk code, Remove Symbols strips formatting symbols, and CodeCipher uses the provided CodeCipher generation scripts with the configured seed. Evaluation of model outputs includes compilation, syntax, and execution checks, and we inspect transformation samples for expected behavior.

For restoration, each obfuscated code is passed through a deobfuscation step to produce a \emph{restored} variant. In the reported experiments, the deobfuscation model is identical to the downstream inference model. The restored code is then passed to the downstream task under the same output constraint as the other protocols (code-only). This design evaluates a same-model restoration pipeline. It does not isolate restoration quality from downstream robustness, so claims about deobfuscation should be read as pipeline-level claims under this setting rather than as model-independent claims about all possible deobfuscators.

\subsection{Inference Protocols}
For each configuration (model, task, language, obfuscation), we run three inference protocols:
\begin{itemize}[leftmargin=*, nosep]
    \item Plain-Inference: downstream inference on the original code.
    \item Obfuscated-Inference: downstream inference on the obfuscated code.
    \item Deobfuscated-Inference: downstream inference on restored code obtained from the obfuscated code.
\end{itemize}

\subsection{Prompting and Decoding}
We adopt a zero-shot prompting strategy with fixed instruction prompts and no demonstrations. Each prompt instructs the model to output \emph{only} the final code wrapped in triple backticks and to match the required entry point signature. Inputs are truncated at 2,000 tokens, and generation is capped at 2,500 new tokens. Decoding is greedy (\texttt{do\_sample=False}) for most open models; for selected configurations, model-specific defaults are applied where required by the implementation. All outputs are post-processed by the same extraction procedure prior to execution-based evaluation: the implementation searches after the \texttt{Result Code} anchor when present, extracts fenced code, strips language labels, skips empty or language-only fences, and falls back to the full output when no fence exists.

\subsection{Metrics}
We evaluate functional correctness using execution-based testing, reporting \textbf{Pass@1} computed against each task's test harness. To isolate the effect of restoration, we define two conditional metrics over the sets of instances passing under each protocol. Let $S_{\mathrm{plain}}$, $S_{\mathrm{obf}}$, and $S_{\mathrm{deobf}}$ denote the sets of instances that pass under Plain-, Obfuscated-, and Deobfuscated-Inference, respectively.

\begin{itemize}[leftmargin=*, nosep]
    \item \textbf{Rescue Rate} measures how often restoration recovers cases broken by obfuscation:
    \[
        \mathrm{RescueRate} =
        \frac{\left| \left(S_{\mathrm{plain}} \setminus S_{\mathrm{obf}}\right) \cap S_{\mathrm{deobf}} \right|}
             {\left| S_{\mathrm{plain}} \setminus S_{\mathrm{obf}} \right|}.
    \]
    \item \textbf{Degradation Rate} measures how often restoration breaks cases that were already robust to obfuscation:
    \[
        \mathrm{DegradationRate} =
        \frac{\left| \left(S_{\mathrm{plain}} \cap S_{\mathrm{obf}}\right) \setminus S_{\mathrm{deobf}} \right|}
             {\left| S_{\mathrm{plain}} \cap S_{\mathrm{obf}} \right|}.
    \]
\end{itemize}

We also report compilation rate, syntax-validity rate, CodeBLEU, edit distance, AST distance, Halstead length, and CodeBERT similarity as supplementary signals. These metrics help characterize surface, structural, and representation changes, but we treat execution as the primary reliability measure.

\subsection{Statistical Analysis}
We primarily report descriptive execution-based rates and paired differences across inference protocols. Where aggregate statistical tests are informative, we treat them as supplementary evidence rather than the sole basis for our conclusions, since the main practical question is whether deobfuscation changes executable outcomes in consistent directions across models and settings.

\section{Results}

\begin{table}[!ht]
    \centering
    \caption{Model performance on code translation (Pass@1). Rescue/Degradation rates are conditional on Plain-Inference success.}
    \label{tab:chain_summary}
    {\footnotesize\emph{Note:} Rescue $=$ $(\text{plain }\checkmark,\ \text{obf }\times,\ \text{deobf }\checkmark)\,/\,(\text{plain }\checkmark,\ \text{obf }\times)$; \\ Degrade $=$ $(\text{plain }\checkmark,\ \text{obf }\checkmark,\ \text{deobf }\times)\,/\,(\text{plain }\checkmark,\ \text{obf }\checkmark)$.\par}
    \vspace{1pt}
    \resizebox{\linewidth}{!}{
    % \midrule
    \begin{tabular}{@{}lrrrrr@{}}
    \toprule
    & \multicolumn{3}{c}{\textbf{Inference}} & \multicolumn{2}{c}{\textbf{Analysis}} \\
    \cmidrule(lr){2-4} \cmidrule(lr){5-6}
    \textbf{Model} & \textbf{Plain} & \textbf{Obf} & \makecell{\textbf{Deobf} \\ \textbf{-Only}} & \textbf{Rescue}$\uparrow$ & \textbf{Degrade}$\downarrow$ \\
    \midrule
CodeLlama-7B & 57.1\% & 48.9\%  &  \makecell{41.0\%\\43.5\%} & 32.2\% & 33.5\% \\
CodeLlama-70B & 16.8\% & 15.2\%  &  \makecell{3.8\%\\3.2\%} & 6.8\% & 95.2\% \\
DeepSeek-Coder-V2 & 76.5\% & 73.0\%  &  \makecell{60.7\%\\74.7\%}  & 55.5\% & 29.5\% \\
DeepSeek-R1-Qwen-14B & 53.9\% & 67.2\%  &  \makecell{56.5\%\\71.3\%}  & 53.7\% & 32.6\% \\
Qwen2.5-Coder-14B & 71.8\% & 59.6\%  &  \makecell{61.8\%\\54.6\%} & 58.3\% & 27.6\% \\
Qwen3-Coder-30B & 88.3\% & 90.5\%  &  \makecell{88.3\%\\71.2\%} & 64.0\% & 5.7\% \\
GPT-4.1 & 92.9\% & 90.0\%  &  \makecell{86.8\%\\94.5\%} & 69.8\% & 8.4\% \\
    \bottomrule
    \end{tabular}
    }
\end{table}

\begin{figure}[t]
    \centering
    \includegraphics[width=\linewidth]{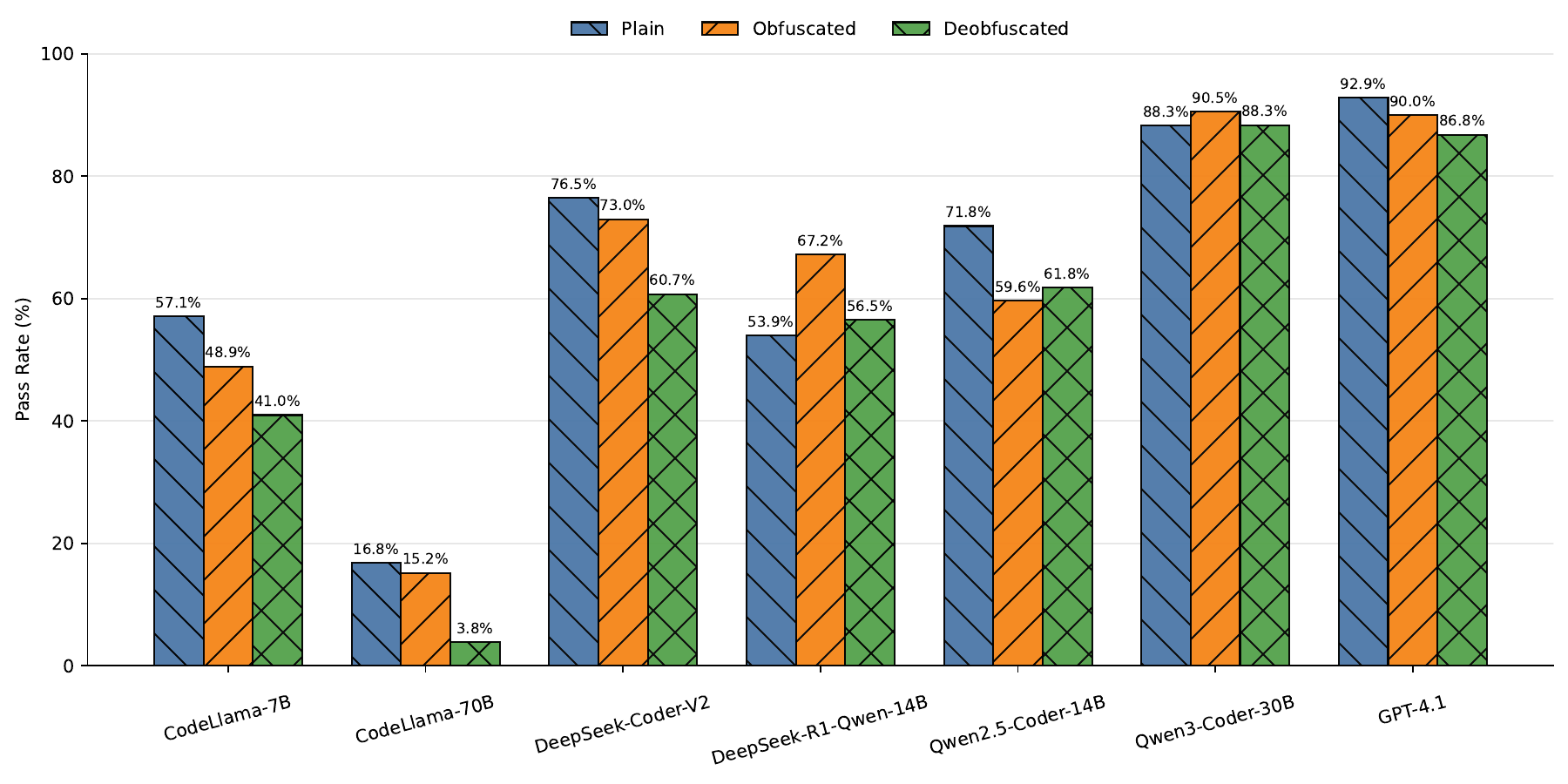}
    \caption{Model-level comparison across the three translation settings. For each model, the bars report Pass@1 under \textbf{Plain}, \textbf{Obfuscated}, and \textbf{Deobfuscated} inference. The figure highlights a central pattern: direct processing of protected code often remains competitive with plain inference, while the restoration stage introduces additional losses for most models.}
    \label{fig:model_comparison}
\end{figure}

Figure~\ref{fig:model_comparison} summarizes the overall empirical results of our study. Two models, GPT-4.1 and Qwen3-Coder-30B, remain close to their plain baselines even on protected inputs, which indicates that higher-capability models can often recover semantics without an explicit restoration stage. At the same time, the deobfuscated bars show that same-model restoration is not a monotonic improvement over direct protected-code inference. For six of the seven models, the extra stage either reduces accuracy or leaves it unchanged.

\begin{table}[htbp]
  \centering
  \caption{Metric disagreement across measurement families. Mean~$\Delta$ = Deobfuscated$-$Plain. \textbf{Bold} indicates Favorable~Rate(Fav.)~$>50\%$, i.e., deobfuscation helps in the majority of cases under that metric.}
  \label{tab:metric_aagreement}
    \resizebox{\linewidth}{!}{
  \begin{tabular}{@{}lllrr@{}}
    \toprule
    \textbf{Task} & \textbf{Family} & \textbf{Metric} & \textbf{$\Delta$} & \textbf{Fav.} \\
    \midrule
    \multirow{8}{*}{Translation} & \multirow{2}{*}{\textit{Exec}} & Pass@1 & -0.0640 & 19.1\% \\
     &  & Compilation Rate & -0.0241 & 25.7\% \\
    \cmidrule(lr){2-5}
     & \multirow{5}{*}{\textit{Static}} & CodeBLEU & -0.0001 & 42.6\% \\
     &  & EditDist & -0.0332 & 34.6\% \\
     &  & AST Dist & +5.42e+03 & 40.4\% \\
     &  & Halstead & +0.41 & 39.7\% \\
     &  & Syntax Valid & -0.0815 & 20.6\% \\
    \cmidrule(lr){2-5}
     & \multirow{1}{*}{\textit{Vector}} & $Sim_{CodeBERT}$ & \textbf{-0.0027} & \textbf{54.4\%} \\
    \midrule
    \multirow{8}{*}{Completion} & \multirow{2}{*}{\textit{Exec}} & Pass@1 & -0.1054 & 42.9\% \\
     &  & Compilation Rate & -0.0854 & 28.6\% \\
    \cmidrule(lr){2-5}
     & \multirow{5}{*}{\textit{Static}} & CodeBLEU & \textbf{+0.0503} & \textbf{71.4\%} \\
     &  & EditDist & \textbf{+0.0967} & \textbf{57.1\%} \\
     &  & AST Dist & -3.83e+04 & 28.6\% \\
     &  & Halstead & -24.41 & 42.9\% \\
     &  & Syntax Valid & -0.0889 & 28.6\% \\
    \cmidrule(lr){2-5}
     & \multirow{1}{*}{\textit{Vector}} & $Sim_{CodeBERT}$ & \textbf{+0.0023} & \textbf{71.4\%} \\
    \bottomrule
  \end{tabular}%
  }
\end{table}

\begin{figure}[htbp]
    \centering
    \includegraphics[width=\linewidth]{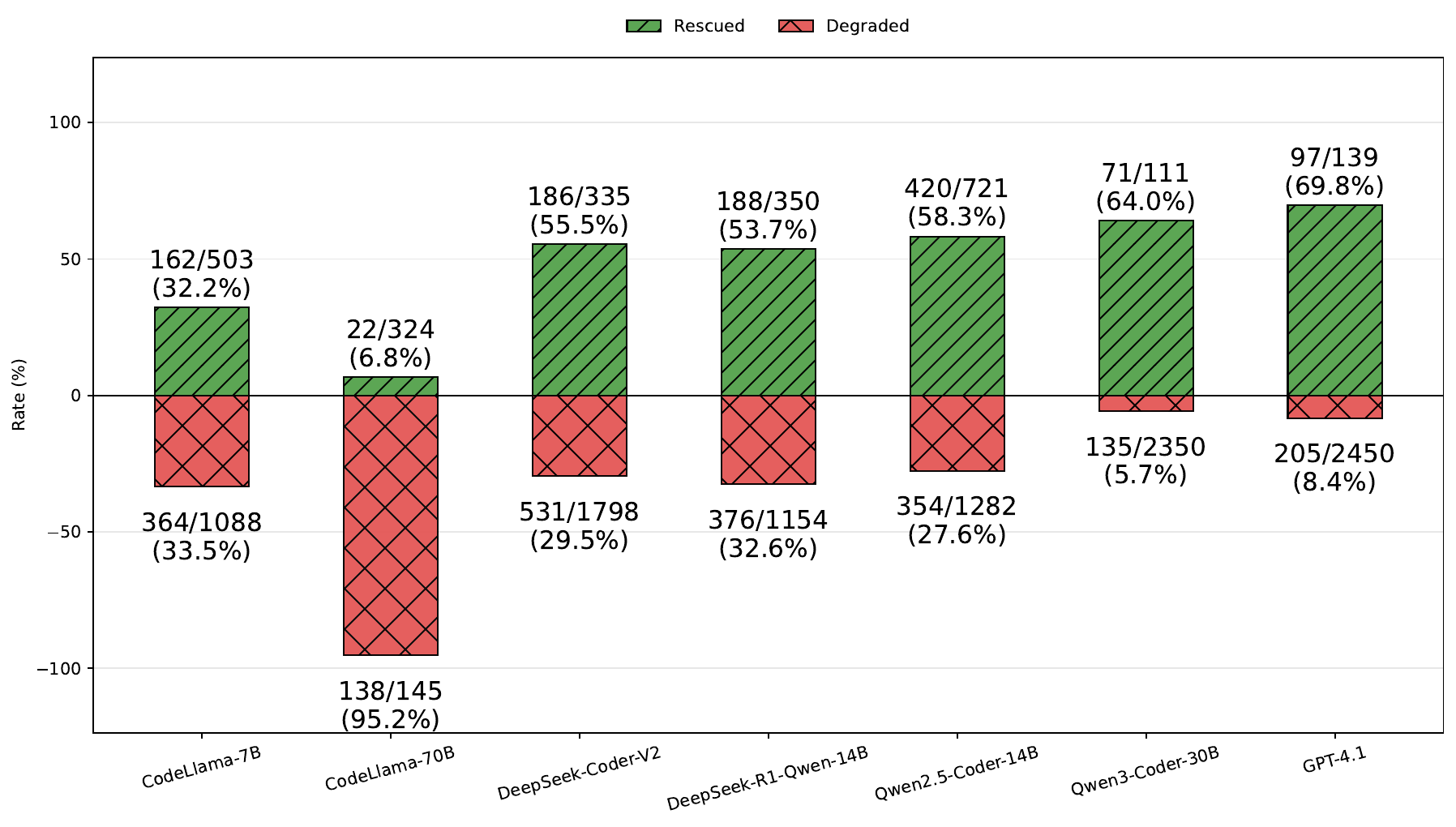}
    \caption{Rescue versus degradation trade-off by model. The x-axis lists models and the y-axis shows conditional rates in percent. Green bars are \textbf{Rescued} rates, defined as obf-fail $\rightarrow$ after-pass among plain-success cases; red bars are \textbf{Degraded} rates, defined as obf-pass $\rightarrow$ after-fail among plain-success cases. The count labels above each bar report numerator/denominator and percentage. Larger degradation than rescue implies net harm from restoration.}
    \label{fig:rescue_degrade}
\end{figure}

\begin{figure}[ht]
    \centering
    \includegraphics[width=\linewidth]{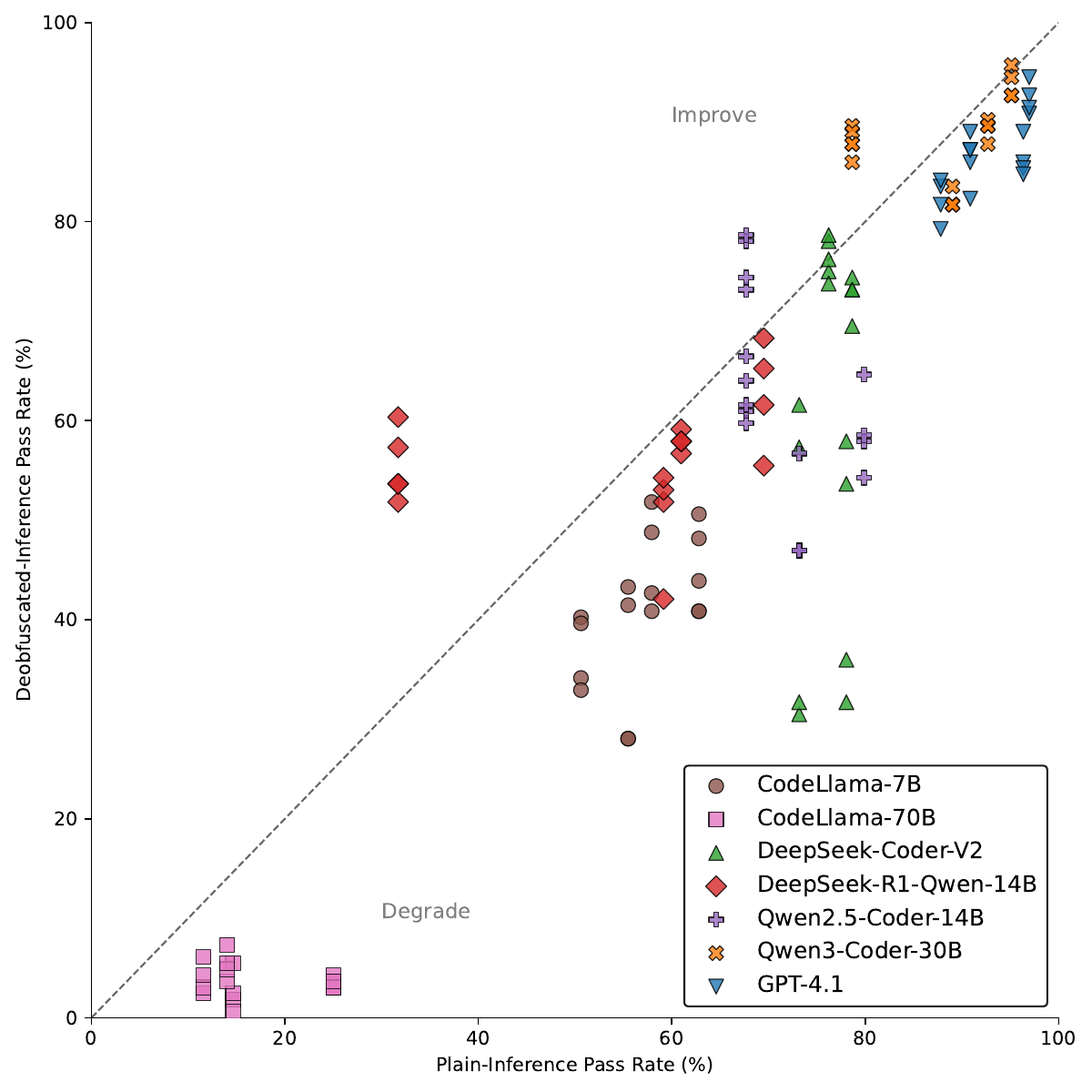}
    \caption{Plain versus Deobfuscated pass rate. Each point is one model against one obfuscator. The x-axis is average Pass@1 under \textbf{Plain} inference, and the y-axis is average Pass@1 under \textbf{Deobfuscated}. Points near the diagonal indicate little net change; points below it lose accuracy after restoration. High plain capability does not guarantee positive gain from restoration, but very low plain capability almost always predicts degradation after restoration.}
    \label{fig:plain_vs_after}
\end{figure}

\section{Analysis}
\label{sec:analysis}

We present our empirical findings organized around the four research questions defined in Section~\ref{sec:rqs}. For each RQ, we first report quantitative results, then provide qualitative analysis with concrete examples, and finally discuss the implications for practice.

\subsection{RQ1: Direct Processing vs.\ Deobfuscation}
\label{sec:rq1}

\subsubsection{Quantitative Results}

Table~\ref{tab:chain_summary} compares three inference settings: Plain, Obfuscated, and Deobfuscated. We observe two distinct patterns across the seven models. First, the strongest models retain near-plain performance under obfuscated inference, indicating that they can often operate directly on protected inputs. Second, under our same-model restoration design, deobfuscation reduces accuracy more often than it improves it.

GPT-4.1 and Qwen3-Coder-30B retain high accuracy on protected code, reaching 90.0\% and 90.5\% Pass@1 under obfuscated inference, respectively. These values are only about 3 percentage points below their plain-code performance (92.9\% and 88.3\%), suggesting that stronger models can often extract semantics directly from protected inputs without an explicit restoration step. The obfuscated-to-plain ratio is 96.9\% for GPT-4.1 and 97.1\% for Qwen3-Coder-30B, which is close to parity.

By contrast, same-model restoration more often reduces than improves performance. Comparing the Obfuscated and Deobfuscated columns, six of the seven models lose accuracy after restoration. The only exception is Qwen2.5-Coder-14B, which gains 2.2 percentage points (59.6\% $\rightarrow$ 61.8\%). CodeLlama-70B shows the largest decline, dropping from 15.2\% to 3.8\%, which corresponds to a 74.8\% relative reduction. DeepSeek-Coder-V2 also degrades substantially, falling from 73.0\% to 60.7\% (16.8\%$\downarrow$).

Figure~\ref{fig:plain_vs_after} visualizes this pattern at the item level: each point represents one model against one obfuscator. Points near the diagonal indicate little net change; points below the diagonal lose accuracy after restoration. The majority of points (78\%) fall below the diagonal, confirming that same-model restoration reduces accuracy in most cases. The gap is smallest for higher-capability models (GPT-4.1, Qwen3-Coder-30B) and largest for lower-capability ones (CodeLlama-70B).

Despite having 10$\times$ more parameters, CodeLlama-70B achieves only 16.8\% plain pass rate compared to 7B's 57.1\%. In our manual inspection, 70B often produces more verbose outputs with multiple alternative solutions, explanations, and overly complex implementations, whereas 7B more often emits short direct solutions. We treat this as a hypothesis about output behavior rather than a definitive causal explanation: parameter count alone does not guarantee better performance when generation style is poorly aligned with the task constraints.

\subsubsection{Qualitative Analysis: Why Does Deobfuscation Harm Lower-Capability Models?}

To understand the mechanism behind deobfuscation-induced degradation, we manually inspected failure cases from lower-capability models. Three recurring patterns emerged. First, restoration can introduce semantic distortion even when the output remains syntactically valid. In one Java example that computes array sums, the deobfuscator incorrectly renames a loop counter and changes the iteration bound from \texttt{arr.length} to \texttt{arr.length - 1}, producing an off-by-one error. Second, restoration sometimes introduces unnatural or non-idiomatic constructs that appear to distract the downstream model. In one JavaScript case, a simple \texttt{map} expression is rewritten as a much longer \texttt{for}-loop, tripling the token count and obscuring the main logic. Third, restoration can remove structural cues that the model had learned to exploit in protected code. Distinctive placeholder names such as \texttt{var\_1} and \texttt{var\_2} may function as anchors for attention, whereas the restored version replaces them with generic names such as \texttt{x} and \texttt{y}.

The same failure classes still appear for GPT-4.1 and Qwen3-Coder-30B, but less frequently and with much smaller aggregate effect. This pattern suggests that higher-capability models are less sensitive to restoration artifacts because their internal representations are more stable under input perturbation.

\subsubsection{Implications for Practice}

The answer to RQ1 has direct implications for pipeline design. For models above roughly 80\% plain accuracy, direct processing of protected code is the safest default because same-model restoration adds little in our results and can still introduce new errors. For models in the 50--80\% range, an adaptive strategy is more plausible: attempt obfuscated inference first and invoke restoration only if the first pass fails or is flagged by a verifier. For models below 50\% plain accuracy, restoration should generally be avoided because these models appear least able to absorb restoration artifacts and their degradation rates exceed their rescue gains. Overall, RQ1 indicates that higher-capability models can operate directly on protected code with limited loss, whereas restoration is risky for weaker models and only occasionally beneficial in the middle band.

\subsection{RQ2: The Rescue--Degradation Trade-off}
\label{sec:rq2}

\subsubsection{Quantitative Results}

Deobfuscation has two opposing effects: it \textit{rescues} cases that failed under obfuscated inference, but \textit{degrades} cases that previously succeeded. Figure~\ref{fig:rescue_degrade} quantifies both effects using conditional rates (defined in Section~\ref{sec:rqs}).

Rescue rates range from 32.2\% to 69.8\%. Higher-capability models recover most obfuscation-induced failures: GPT-4.1 reaches 69.8\% rescue and Qwen3-Coder-30B reaches 64.0\%. Mid-tier models also recover a substantial fraction of failures, with DeepSeek-Coder-V2 at 55.5\% and Qwen2.5-Coder-14B at 58.3\%. Lower-capability models recover far fewer cases; CodeLlama-70B manages only 6.8\% rescue, which means it fails to recover more than 93\% of obfuscation-induced failures.

Degradation rates vary even more strongly. Qwen3-Coder-30B and GPT-4.1 have the lowest degradation rates at 5.7\% and 8.4\%, so they rarely break cases that were already working. At the other extreme, CodeLlama-70B degrades 95.2\% over its successful cases, meaning that restoration severely harms this model. DeepSeek-R1-Qwen-14B shows the next-highest degradation rate at 32.6\%, which suggests that reasoning-oriented training alone does not make a model robust to restoration artifacts.

We summarize the trade-off through the net effect, defined as rescue minus degradation. GPT-4.1 has a strongly positive net effect (+61.4 percentage points), so rescue substantially outweighs collateral damage. CodeLlama-70B, by contrast, has a deeply negative net effect ($-$88.4 points), where degradation overwhelmingly dominates. This asymmetry explains why deobfuscation helps some models while harming others.

Across the seven models, three have positive net effects (GPT-4.1, Qwen3-Coder-30B, Qwen2.5-Coder-14B), three have negative net effects (DeepSeek-R1-Qwen-14B, DeepSeek-Coder-V2, CodeLlama-7B), and CodeLlama-70B shows the most severe negative effect. Notably, all models with positive net effects have plain pass rates above 70\%, which is consistent with a capability threshold for beneficial deobfuscation.

\subsubsection{Case Study: When Rescue Succeeds vs.\ Fails}

To illustrate the rescue--degradation dynamic, we examine two contrasting cases from our dataset. In a rescued case, a Java method implementing binary search was obfuscated through identifier renaming, replacing \texttt{low}, \texttt{high}, and \texttt{mid} with \texttt{a}, \texttt{b}, and \texttt{c}. GPT-4.1 failed on the obfuscated input because it set the loop condition incorrectly, but once deobfuscation restored the original names, the model generated the correct algorithm. This is a clean instance of rescue: the missing semantic cues were restored and the downstream model benefited.

In a degraded case, a Python factorial function was obfuscated with dead-code injection via an unreachable \texttt{print} statement. CodeLlama-70B correctly ignored that noise and produced the right answer on the protected input. After deobfuscation removed the dead code, however, the same model misread the simplified structure and generated an incorrect recursive solution. This is a clear degradation case: the model had learned to cope with the protected representation, and restoration disrupted that coping behavior.

These cases support our quantitative findings: rescue succeeds when deobfuscation restores lost semantics, but degradation occurs when deobfuscation disrupts learned coping strategies.

\subsubsection{Implications for Adaptive Pipelines}

The coexistence of rescue and degradation motivates an adaptive pipeline. A practical policy is to attempt obfuscated inference first, skip deobfuscation when that first pass succeeds, and use restoration only as a fallback when the direct path fails. The thresholds should remain model-aware: for stronger models with degradation below roughly 10\%, restoration is less risky, whereas for weaker models with degradation above 30\%, restoration should generally be avoided. We return to this conditional strategy in Section~\ref{sec:discussion}. Taken together, the RQ2 results show that rescue and degradation coexist rather than cancel cleanly, and that the value of restoration depends primarily on model capability.

\subsection{RQ3: Boundary Conditions}
\label{sec:rq3}

We analyze how model capability, source language, and protection method jointly influence the rescue--degradation trade-off. These factors are relevant for designing adaptive pipelines that recover failures without introducing unnecessary collateral loss.

\subsubsection{Model Capability is the Primary Factor}
\label{sec:model-capability}

Table~\ref{tab:chain_summary} shows a strong association between plain performance and deobfuscation outcomes. Models with high plain pass rates (GPT-4.1: 92.9\%, Qwen3-Coder-30B: 88.3\%) exhibit low degradation and positive net effects. Models with low plain pass rates (CodeLlama-70B: 16.8\%) exhibit the largest degradation.

Plain pass rate tracks deobfuscation outcome closely: models with stronger plain performance also show lower degradation and more favorable net effects. Plain capability is therefore a useful predictor of whether deobfuscation is likely to improve or reduce downstream accuracy.

Figure~\ref{fig:plain_vs_after} provides additional evidence: models with high plain accuracy cluster near the diagonal (minimal change after deobfuscation), while models with low plain accuracy fall well below the diagonal (substantial degradation). The figure indicates that restoration preserves performance primarily for models that already perform well on plain code.

Table~\ref{tab:by_model_lang} further shows that higher-capability models maintain high deobfuscation pass rates across all languages (GPT-4.1: 82--92\%, Qwen3-Coder-30B: 82--94\%), while lower-capability models remain at very low levels (CodeLlama-70B: 2--5\%).
This consistency across languages reinforces that model capability, not language-specific factors, drives the primary effect.

We observe a practical separation around the 70\% plain pass-rate region: models above this range are more likely to benefit from deobfuscation, while models below it are more likely to suffer net harm. We treat this as a heuristic rather than a hard statistical threshold.

This pattern is reinforced by a supplementary analysis of the most difficult protection slice (\textit{CodeCipher} with downstream inference after deobfuscation). Using a cross-metric robustness score that aggregates static, vector, and execution signals, the ranking is DeepSeek-Coder-V2 (z-avg $=0.792$), GPT-4.1 ($0.678$), Qwen3-Coder-30B ($0.175$), Qwen2.5-Coder-14B ($0.019$), DeepSeek-R1-Qwen-14B ($-0.033$), CodeLlama-7B ($-0.412$), and CodeLlama-70B ($-1.012$). Although this analysis is supplementary, it supports the same main conclusion as the execution-based tables: robustness under restoration is not a monotonic function of nominal model size, and the weakest behavior is concentrated in models that already have poor plain-code capability.

This supplementary ranking is useful because it shows that the model-capability effect is not tied to Pass@1 alone. The same ordering reappears when the comparison includes syntax validity, structural similarity, and embedding-based similarity. That consistency strengthens the interpretation that restoration sensitivity is a model property rather than an artifact of a single metric family.

\subsubsection{Language Effects are Secondary but Consistent}
\label{sec:language-effects}

\begin{table}[htbp]
\centering
\caption{Performance by Programming Language (Pass@1). Pass rates under three conditions (Plain, Obfuscated, Deobfuscated) and change rates (Rescued: obfuscated$\rightarrow$deobfuscated improvement, Degraded: plain$\rightarrow$deobfuscated decline).}
\label{tab:by_language}
\resizebox{\linewidth}{!}{%
    \begin{tabular}{l ccc cc}
    \toprule
    & \multicolumn{3}{c}{\textbf{Pass Rate}} & \multicolumn{2}{c}{\textbf{Change Rate}} \\
    \cmidrule(lr){2-4} \cmidrule(lr){5-6}
    \textbf{Language} & \textbf{Plain} & \textbf{Obfuscated} & \textbf{Deobfuscated} & \textbf{Rescued} $\uparrow$ & \textbf{Degraded} $\downarrow$ \\
    \midrule
C++ & 56.2\% & 53.7\% & 49.5\% & 47.5\% & 28.2\% \\
Go & 59.5\% & 55.7\% & 47.4\% & 41.0\% & 34.4\% \\
Java & 54.1\% & 57.0\% & 53.6\% & 42.4\% & 24.5\% \\
JavaScript & 59.6\% & 55.3\% & 48.1\% & 42.2\% & 31.8\% \\
    \bottomrule
    \end{tabular}%
}
\end{table}

\begin{table}[t]
    \centering
    \caption{Average pass rate by protection method across all seven models. Rows are protection methods and columns show \textbf{Obfuscated} and \textbf{Deobfuscated} pass rates, along with \textbf{Rescued} and \textbf{Degraded} rates. Values are mean Pass@1 percentages aggregated over models on Java language.}
    \label{tab:by_obfuscator}
    \resizebox{\linewidth}{!}{%
    \begin{tabular}{l cc cc}
    \toprule
    & \multicolumn{2}{c}{\textbf{Pass Rate}} & \multicolumn{2}{c}{\textbf{Change Rate}} \\
    \cmidrule(lr){2-3} \cmidrule(lr){4-5}
\textbf{Obfuscation} & \textbf{Obfuscated} & \textbf{Deobfuscated} & \textbf{Rescued} $\uparrow$ & \textbf{Degraded} $\downarrow$ \\
    \midrule
Identifier & 58.2\% & 54.0\% & 50.7\% & 23.4\% \\
Dead Code & 59.9\% & 54.0\% & 42.1\% & 27.1\% \\
Remove Symbols & 56.6\% & 53.2\% & 36.2\% & 23.4\% \\
Random & 56.6\% & 54.3\% & 51.7\% & 23.7\% \\
CodeCipher & 53.7\% & 52.4\% & 31.2\% & 24.6\% \\
    \bottomrule
    \end{tabular}%
    }
\end{table}

\begin{table}[htbp]
\centering
\caption{Exact deobfuscation pass rates by model and source language. Figure~\ref{fig:model_lang_heatmap} visualizes the same matrix to highlight cross-model differences in language sensitivity.}
\label{tab:by_model_lang}
\begin{tabular}{lcccc}
\toprule
\textbf{Model} & \textbf{C++} & \textbf{Go} & \textbf{Java} & \textbf{JavaScript} \\
\midrule
CodeLlama-7B & 36.7\% & 35.2\% & 44.9\% & 46.0\% \\
CodeLlama-70B & 2.6\% & 4.0\% & 3.5\% & 5.3\% \\
DeepSeek-Coder-V2 & 72.6\% & 44.8\% & 76.3\% & 45.3\% \\
DeepSeek-R1-Qwen-14B & 57.9\% & 50.3\% & 55.4\% & 62.7\% \\
Qwen2.5-Coder-14B & 61.6\% & 58.8\% & 74.1\% & 49.4\% \\
Qwen3-Coder-30B & 82.2\% & 93.9\% & 88.0\% & 89.3\% \\
GPT-4.1 & 82.2\% & 92.4\% & 86.3\% & 86.3\% \\
\bottomrule
\end{tabular}
\end{table}

Table~\ref{tab:by_language} shows that rescue rates are relatively consistent across languages (41--48\%), but degradation varies more (24--34\%). Go has the highest degradation (34.4\%) and the lowest deobfuscated pass rate (47.4\%), suggesting that Go code is more sensitive to restoration artifacts. Java shows the most stable behavior after restoration, with the lowest degradation (24.5\%). We interpret these language effects as consistent descriptive trends rather than as the main source of variation.

Two factors may help explain Go's higher sensitivity. First, Go's rigid syntax and type discipline make it less tolerant of restoration artifacts, so even a small off-by-one error or type mismatch can turn into compilation failure. Second, Go relies heavily on idiomatic constructs such as \texttt{range} loops and standard error-handling patterns. A restored program can be semantically close to the original while still becoming sufficiently non-idiomatic to confuse the downstream model.

Operationally, this result suggests additional caution when processing Go code: direct inference on protected code is preferable unless restoration is justified by additional evidence.

\begin{figure}[t]
    \centering
    \includegraphics[width=\linewidth]{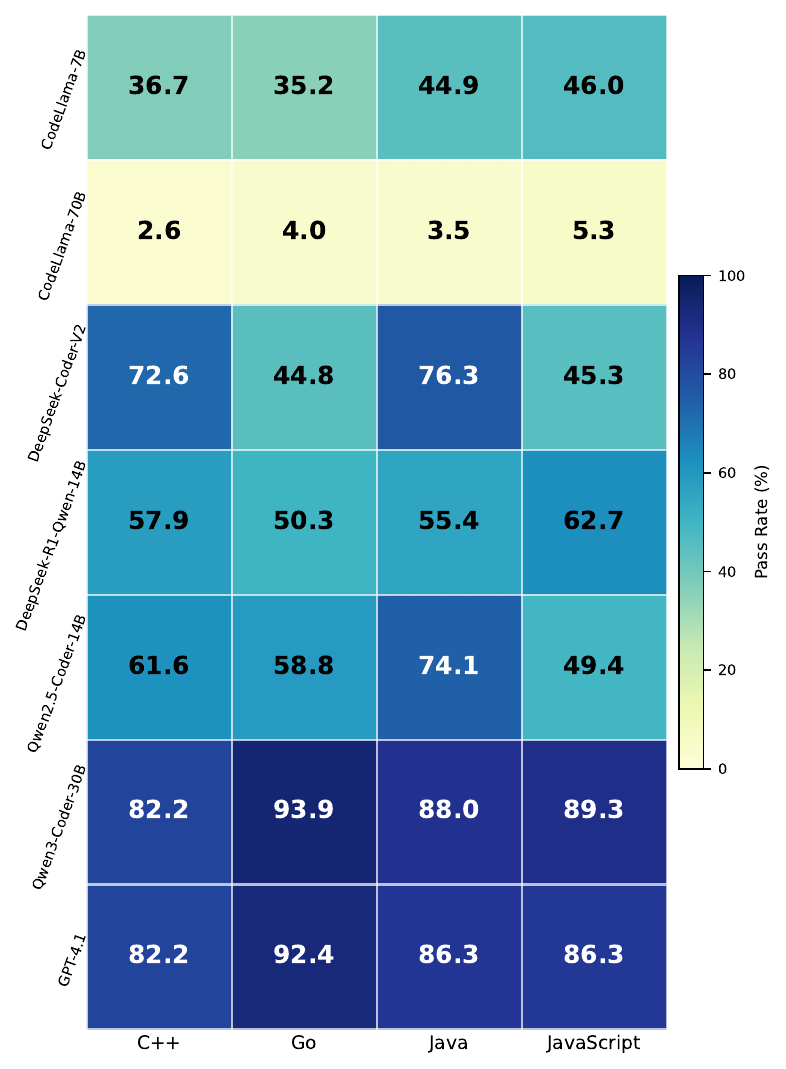}
    \caption{After-deobfuscation Pass@1 by model and source language. The heatmap shows that language effects depend on model capability rather than appearing uniformly across the benchmark. Higher-capability models remain consistently high across all four languages, while mid-tier models exhibit language-specific failure patterns.}
    \label{fig:model_lang_heatmap}
\end{figure}

The aggregate language averages should nevertheless be interpreted cautiously. Figure~\ref{fig:model_lang_heatmap} and Table~\ref{tab:by_model_lang} show that the language effect is not uniform across models. GPT-4.1 and Qwen3-Coder-30B remain consistently strong on all four languages, whereas DeepSeek-Coder-V2 drops sharply on Go and JavaScript while remaining relatively strong on C++ and Java. Qwen2.5-Coder-14B exhibits the opposite asymmetry: Java remains strong, but JavaScript is much less stable.

This cross-model heterogeneity means that source language is best treated as a conditional routing feature rather than as a global policy variable. A team using a high-capability model may standardize on direct inference for all four languages, whereas a team using a mid-tier model may require different defaults for Java-heavy and Go-heavy workloads. Language effects become operationally important only when paired with model capability.

\subsubsection{Obfuscation Method Effects on the Java Slice}
\label{sec:obfuscation-method}

Table~\ref{tab:by_obfuscator} shows a distinct result on the Java slice: CodeCipher, the most aggressive protection method in our set, yields the lowest rescue rate (31.2\%) but moderate degradation (24.6\%). In contrast, simpler methods such as Random and Identifier yield higher rescue rates (51.7\% and 50.7\%).

One possible explanation is transformation consistency: structured, aggressive transformations may produce artifacts that the deobfuscator handles more predictably, whereas simpler transformations can leave subtle inconsistencies that are harder to resolve correctly. CodeCipher applies learned transformations optimized for reversibility, so a deobfuscator trained on similar distributions may invert them more consistently and return structurally coherent code even when some semantics are lost. By contrast, identifier renaming and dead-code injection are applied heuristically, without any explicit optimization for reversibility. The deobfuscator must infer original names and remove injected code, which creates more opportunities for subtle mistakes such as an incorrect variable name or a missed branch.

From a protection-design perspective, this distinction matters. If the goal is to prevent LLM-based deobfuscation, CodeCipher-style learned protection appears more effective than traditional methods. If the goal is to preserve LLM utility on protected code for authorized maintenance or migration, simpler methods may still be preferable because they allow higher rescue rates when deobfuscation is needed. Overall, the RQ3 analysis indicates that model capability remains the dominant factor, while language and obfuscation method act as secondary but still operationally relevant boundary conditions.

\subsection{RQ4: Practical Guidance}
\label{sec:rq4}

Based on our findings, we provide decision rules for pipeline selection in production settings.

\subsubsection{Decision Rules for Pipeline Design}

Model capability should be assessed before any pipeline choice is made. When a model exceeds roughly 80\% pass rate on plain code, direct processing of protected inputs is usually the better default because deobfuscation adds little benefit in our results and still carries some degradation risk. GPT-4.1 and Qwen3-Coder-30B fall into this category.

Models in the 50--80\% range should be treated as routing-sensitive rather than deobfuscation-first. These models show meaningful rescue opportunities, but the aggregate evidence in this paper is not yet strong enough to justify a single fixed deobfuscation policy for all cases. A better default is to attempt obfuscated inference first and reserve deobfuscation for situations where additional evidence suggests the first pass is unreliable, such as failed execution, low confidence, or disagreement with a downstream verifier.

For models below 50\% plain pass rate, deobfuscation should generally be avoided because the degradation risk far exceeds the rescue potential. CodeLlama-70B is the clearest example: deobfuscation reduces its accuracy from 15.2\% to 3.8\%, a 75\% relative drop. Language and obfuscation context still matter around these broad capability bands. Go inputs carry higher degradation risk, so direct inference is preferable unless restoration is supported by additional evidence. CodeCipher-protected inputs tend to have lower rescue rates but more predictable behavior, which may still be acceptable in settings where predictability matters more than maximal accuracy.

These rules can be implemented as an obfuscated-first policy. A deployment would run direct inference on the protected input, check the output with tests or a lightweight verifier, and accept it when the check passes. If the direct path fails or confidence is low, the pipeline can invoke restoration and accept the restored-path output only if it passes the same check. If neither path passes, the item should be escalated. The main open challenge is the trigger, since many workflows lack complete tests, calibrated confidence, or cheap semantic verification.

\subsubsection{Metric Selection Guidance}

Table~\ref{tab:metric_aagreement} shows that static metrics (CodeBLEU, EditDist) and vector similarity ($Sim_{\text{CodeBERT}}$) do not always align with execution outcomes. In the Translation task, only 19.1\% of cases show favorable Pass@1 changes, but 42.6\% show favorable CodeBLEU changes. This divergence means a paper based on only one static metric would overstate certainty.

We therefore recommend using Pass@1, or an equivalent execution-based measure, as the primary decision criterion because it directly captures functional correctness. Compilation rate is a useful secondary signal for syntax-level sanity checks, especially when restoration introduces malformed outputs. Static metrics such as CodeBLEU and EditDist, along with vector similarity, remain informative as descriptive context, but they should not be treated as interchangeable with execution outcomes.

We also checked simple complexity measures over saved generated outputs. Relative to Plain-Inference outputs, Deobfuscated-Inference outputs are slightly longer in mean NLOC (translation: 7.60$\rightarrow$8.50; completion: 13.80$\rightarrow$16.84), while estimated AST-based cyclomatic complexity decreases (translation: 3.31$\rightarrow$3.10; completion: 3.65$\rightarrow$3.04). This suggests that restoration often rewrites outputs into longer forms without increasing branch complexity. These measures complement the Halstead and AST-distance results in Table~\ref{tab:metric_aagreement}, but they do not replace execution-based correctness.

\subsubsection{Cost--Benefit Analysis}

Deobfuscation adds computational overhead: one additional inference pass plus deobfuscation runtime. For higher-capability models, skipping deobfuscation saves 30--50\% latency with little accuracy loss. For lower-capability models, attempting deobfuscation consumes additional compute while reducing accuracy. For mid-tier models, the cost of deobfuscation is justified only when routing is informed by some item-level signal; absent that signal, direct inference on protected code remains the safer default.

\begin{table}[t]
    \centering
    \caption{Conditional success after restoration. $P(\mathrm{After}\mid\mathrm{Obf}=1)$ measures how often a model remains correct after restoration given that it already succeeded directly on protected code. $P(\mathrm{After}\mid\mathrm{Obf}=0)$ measures how often restoration recovers previously failed protected-code cases.}
    \label{tab:conditional_success}
    \resizebox{\linewidth}{!}{%
    \begin{tabular}{lcc}
    \toprule
    \textbf{Model} & \boldmath$P(\mathrm{After}\mid\mathrm{Obf}=1)$ & \boldmath$P(\mathrm{After}\mid\mathrm{Obf}=0)$ \\
    \midrule
    CodeLlama-7B & 66.3\% & 32.8\% \\
    CodeLlama-70B & 4.6\% & 6.7\% \\
    DeepSeek-Coder-V2 & 69.9\% & 55.9\% \\
    DeepSeek-R1-Qwen-14B & 66.2\% & 53.7\% \\
    Qwen2.5-Coder-14B & 71.9\% & 57.8\% \\
    Qwen3-Coder-30B & 94.3\% & 62.5\% \\
    GPT-4.1 & 91.6\% & 76.2\% \\
    \bottomrule
    \end{tabular}
    }
\end{table}

Table~\ref{tab:conditional_success} provides a compact item-level view of the routing problem. The left conditional probability shows how often restoration preserves success on cases already solved directly from protected code; the right one shows how often restoration recovers failures. Higher-capability and lower-capability models differ on both axes. GPT-4.1 and Qwen3-Coder-30B preserve more than 90\% of previously successful protected-code cases while also recovering a substantial fraction of failures. CodeLlama-70B performs poorly on both dimensions: restoration rarely recovers failures and almost never preserves prior success.

The current item-difficulty analysis points in the same direction. For medium-difficulty items, after-deobfuscation pass rates remain high for GPT-4.1 (88.8\%) and Qwen3-Coder-30B (90.5\%), but drop sharply for CodeLlama-7B (41.9\%) and CodeLlama-70B (3.9\%). For hard items, all models remain near zero even after restoration. Restoration therefore does not create a broadly easier regime; instead, it mostly redistributes error across already contested cases. This is the setting in which verifier-backed or confidence-aware routing is more appropriate than a single static preprocessing policy. Overall, the RQ4 evidence supports model-aware and context-aware pipeline selection: stronger models usually skip deobfuscation, weaker models should avoid it, and mid-tier models benefit most from obfuscated-first routing with selective restoration triggered by additional evidence.

\section{Implications}
\label{sec:discussion}

In this section, we discuss broader implications of our findings and the limitations of our study.

\subsection{Implications for LLM Robustness}

Our results challenge the common assumption that deobfuscation uniformly improves LLM performance on protected code. Instead, the effect depends strongly on model capability: same-model restoration is more favorable for higher-capability models and unfavorable for lower-capability ones. One implication is that models differ in their sensitivity to representation changes. The mismatch between rescue and net gain suggests that higher-capability models have internal representations that better tolerate perturbations, whereas lower-capability models are more brittle. This interpretation is consistent with recent work on adversarial robustness in vision and NLP, where stronger models often show greater resistance to input perturbations~\cite{hendrycks2020pretrained}.

Another implication is that models appear to learn coping strategies for protected inputs, and those strategies can be disrupted by restoration. A model may, for example, learn to ignore certain token patterns in obfuscated code, while those same patterns cease to be useful once deobfuscation rewrites the surrounding cues. Understanding these coping strategies would be a productive direction for future work because it bears directly on why restoration helps some models yet harms others.

\subsection{Reliability Engineering Interpretation}

Our results are best interpreted as a study of \emph{pipeline reliability} rather than deobfuscation quality in isolation. A restoration stage is justified only if it improves the probability of correct downstream behavior without introducing a disproportionate number of new failures. The rescue--degradation formulation makes that trade-off explicit by separating recovery from collateral damage, which is central when evaluating whether an additional transformation should be part of a production workflow.

This framing also clarifies why execution-based metrics are central. Static similarity can show that restoration makes code look closer to a natural reference, yet the key question is whether the translated artifact executes correctly. In our setting, restoration often improves representation-level properties without yielding a corresponding reliability gain. Protected-code pipelines should therefore be evaluated end-to-end, where each additional transformation is justified by measurable reduction in downstream failure risk rather than by readability or textual similarity alone.

More broadly, the paper highlights a design pattern that should generalize beyond this benchmark: for AI-assisted engineering systems, preprocessing stages that seem semantically benign can still reduce operational reliability if they perturb the cues a model has learned to exploit. Similar concerns are likely to arise in workflows involving summarization, repair, security review, or migration on transformed artifacts. The relevant question is therefore not simply whether a model understands obfuscation, but whether a full pipeline preserves dependable behavior under realistic transformations and fallbacks.

\subsection{Implications for Operational Deployment}

The results also suggest a deployment lesson: protected-code pipelines should be \emph{observable decision systems} rather than static preprocessing chains. Operators care whether restoration increases correct downstream actions on the current workload, not whether it looks plausible in isolation. A production pipeline that always restores first hides the signal needed to determine whether restoration helps or hurts.

At minimum, a deployed system should retain telemetry to compare direct and restored paths at the slice level. The key transition patterns are recovery cases, collateral failures, and persistent failures. Recovery cases justify the extra preprocessing cost; collateral failures reveal unsafe dependence on restoration; persistent failures point to model choice, prompting, or verification rather than preprocessing.

This perspective changes fallback design. A fixed rule such as ``always deobfuscate first'' or ``never deobfuscate'' is poorly matched to the heterogeneous behavior across models and languages. More robust deployments should treat restoration as one candidate action in a monitored policy, triggered by evidence such as execution failure, low-confidence self-checks, or verifier disagreement. Because restoration becomes part of the trusted workflow, evaluations should report not only average accuracy but also preservation of already-correct cases and stability under fallback.

\subsection{Implications for Obfuscation Design}

Our findings about CodeCipher also have implications for obfuscation design. CodeCipher achieves strong protection in the sense of low rescue rates, yet its degradation remains moderate rather than catastrophic. That combination may be attractive in high-assurance settings where predictable behavior matters more than maximal obstruction. Simpler methods such as identifier renaming provide weaker protection, but they also preserve more utility when authorized deobfuscation is required.

If obfuscation must be reversible for debugging, migration, or maintenance, then explicit reversibility constraints may improve deobfuscation reliability. CodeCipher's learned approach appears to optimize for reversibility implicitly; making that objective more explicit could yield better trade-offs between protection strength and downstream utility.

\subsection{Limitations}

We acknowledge several limitations of our study. First, the evaluation relies on HumanEval-style function-level tasks, which do not represent the full range of code-understanding scenarios encountered in real codebases with dependencies, build systems, long contexts, and stronger idiomatic variation. The benchmark is useful because it provides executable tests and controlled transformations, but repository-scale studies are needed before making broader deployment claims. Extending the analysis to larger benchmarks and real repositories would improve external validity.

Second, although we report both translation and completion, the paper is primarily a translation study. Completion results are included as a comparison point, but they are not developed to the same depth as the translation analysis in the main text.

Third, we evaluate only five obfuscation methods even though the broader design space is much larger and includes techniques such as control-flow flattening and virtualization. Our selection covers several major categories, but the conclusions should not be read as universal across all obfuscation strategies.

Fourth, all experiments use zero-shot inference. Fine-tuned models, few-shot prompting, or training directly on obfuscated code may produce different robustness profiles, and we leave those variants to future work.

Fifth, the practical guidance for conditional deobfuscation should be interpreted cautiously because the present paper provides model-level and conditional item-level evidence rather than a fully implemented adaptive routing policy. A deployed policy also requires reliable triggers, test availability, confidence calibration, semantic verification of restored code, model- and language-specific thresholds, and a fallback path when checks are unavailable.

Finally, our conclusions depend in part on the quality of the deobfuscator. In the reported experiments, restoration is performed by the same model used for downstream inference, so restoration quality and downstream robustness are coupled. A fixed strong or tool-based deobfuscator could achieve higher rescue with lower degradation, and would better isolate restoration quality from downstream robustness. We therefore interpret the results as evidence about the same-model restoration pipeline studied here, not as a model-independent claim about every possible deobfuscator.

\section{Threats to Validity}
\label{sec:threats}

\subsection{Construct Validity}

Our primary construct is Pass@1, which measures functional correctness via test execution. While this is the standard metric for code generation, it may not capture all aspects of code quality (e.g., efficiency, readability). We supplement with static metrics (CodeBLEU, EditDist) to provide additional context, but these also have limitations.

Our code extraction protocol is uniform across models, but it can still interact with model style. Models that emit multiple alternatives or long explanations may be disadvantaged by a code-only scoring protocol even when a correct solution appears later in the response. We therefore treat output-style explanations, including the CodeLlama-70B discussion, as hypotheses rather than causal claims.

The definitions of rescue rate and degradation rate are conditional probabilities, which require careful interpretation. We define these rates conditioned on Plain-Inference success to isolate the effect of deobfuscation from baseline capability. Alternative definitions (e.g., conditioning on Obfuscated-Inference outcomes) would yield different numerical values but should preserve the relative ordering of models.

\subsection{Internal Validity}

A threat to internal validity is the potential confounding between model capability and other factors (e.g., training data, architecture). We control for this by analyzing multiple models across different families (CodeLlama, Qwen, DeepSeek, GPT), but cannot fully rule out confounding. Even so, the descriptive pattern is clear: stronger plain performance is associated with lower degradation and more favorable deobfuscation outcomes.

Another threat is the quality of obfuscation and deobfuscation implementations. We implement deterministic static transformations, use the CodeCipher generation scripts for learned protection, and evaluate outputs through syntax, compilation, and execution checks. Even so, bugs or suboptimal implementations could affect results, and we do not independently prove semantic equivalence for every restored source.

\subsection{External Validity}

Our study focuses on code translation from C++, Go, Java, and JavaScript to Python. Results may not generalize to other language pairs or tasks (e.g., code completion, summarization). We expect similar patterns for tasks requiring deep semantic understanding, but surface-level tasks (e.g., syntax checking) may show different effects.

The models studied were released between 2023 and 2025. Newer models with improved capabilities may exhibit different robustness characteristics. However, the fundamental trade-off between rescue and degradation should persist as long as deobfuscation introduces some perturbations.

\subsection{Conclusion Validity}

Our strongest claims are descriptive rather than purely inferential: we rely primarily on consistent differences in execution-based outcomes across models, languages, and protection settings. The large number of benchmark instances per evaluation slice reduces noise, but some finer-grained conclusions---especially around qualitative failure taxonomy and adaptive routing---should still be interpreted as suggestive rather than definitive.

\section{Conclusion}
\label{sec:conclusion}

We presented an empirical study of code LLM robustness on protected code, examining the trade-offs between direct processing and deobfuscation. Our central finding is that deobfuscation yields mixed effects: it recovers performance on inputs that fail in obfuscated form, yet simultaneously degrades performance on inputs that would otherwise succeed. The net outcome is strongly mediated by model capability, with a meaningful practical threshold emerging around the 70\% plain pass-rate region. Higher-capability models such as GPT-4.1 and Qwen3-Coder-30B sustain over 90\% accuracy on protected code and gain little from deobfuscation, while lower-capability models like CodeLlama-70B are actively harmed by it. These findings suggest a simple model-aware principle for practitioners: assess capability first, then select a pipeline or routing strategy accordingly.

% Higher-capability models such as GPT-4.1 and Qwen3-Coder-30B can often skip deobfuscation altogether. They maintain more than 90\% accuracy on protected code and experience minimal degradation, so direct processing is both effective and more efficient, saving roughly 30--50\% latency. Mid-tier models show meaningful rescue opportunities, but the safer default is still obfuscated-first inference with selective restoration triggered by additional evidence that the first pass is unreliable. Lower-capability models should generally avoid deobfuscation entirely. CodeLlama-70B, for example, suffers 95\% degradation, so restoration almost always reduces accuracy. Language and method effects remain secondary: Go is more sensitive than Java, and CodeCipher produces more predictable but less rescuable outcomes than simpler methods.

% These findings provide guidance for practitioners working on legacy system migration, code auditing, and other scenarios involving protected code. Pipeline design should be model-aware. We recommend assessing model capability first and then selecting a pipeline or routing strategy accordingly.

Future work should evaluate fixed strong and tool-based deobfuscators, repository-scale workloads, broader transformations such as control-flow flattening and virtualization, and fully implemented routing policies with verifier-backed triggers. More broadly, we call for deeper investigation into why some models are more sensitive to restoration-induced input perturbations.

\noindent\textbf{Artifact Availability.} An anonymized artifact package, including code, generated outputs, and per-item result files, is available at \url{https://anonymous.4open.science/r/deobf-9331/}.

\bibliographystyle{IEEEtran}
\bibliography{references}

@inproceedings{hendrycks2020pretrained,
  title={Pretrained transformers improve out-of-distribution robustness},
  author={Hendrycks, Dan and Liu, Xiaoyuan and Wallace, Eric and Dziedzic, Adam and Krishnan, Rishabh and Song, Dawn},
  booktitle={Proceedings of the 58th annual meeting of the association for computational linguistics},
  pages={2744--2751},
  year={2020}
}

@inproceedings{udupa2005deobfuscation,
  title={Deobfuscation: Reverse engineering obfuscated code},
  author={Udupa, Sharath K and Debray, Saumya K and Madou, Matias},
  booktitle={12th Working Conference on Reverse Engineering (WCRE'05)},
  pages={10--pp},
  year={2005},
  organization={IEEE}
}

@manual{proguard_manual,
  title        = {ProGuard Manual},
  author       = {{Guardsquare}},
  organization = {Guardsquare},
  year         = {2026},
  url          = {https://www.guardsquare.com/manual/home},
  note         = {Accessed: 2026-04-22}
}

@misc{collberg1997taxonomy,
  title={A taxonomy of obfuscating transformations},
  author={Collberg, Christian and Thomborson, Clark and Low, Douglas},
  year={1997},
  publisher={Technical Report 148, Department of Computer Science, University of Auckland}
}

@book{nagra2009surreptitious,
  title={Surreptitious software: obfuscation, watermarking, and tamperproofing for software protection},
  author={Nagra, Jasvir and Collberg, Christian},
  year={2009},
  publisher={Pearson Education}
}

@article{schrittwieser2016softwareobfuscation,
  author  = {Sebastian Schrittwieser and Stefan Katzenbeisser and Johannes Kinder and Georg Merzdovnik and Edgar Weippl},
  title   = {Protecting Software through Obfuscation: Can It Keep Pace with Progress in Code Analysis?},
  journal = {ACM Computing Surveys},
  volume  = {49},
  number  = {1},
  pages   = {4:1--4:37},
  year    = {2016},
  doi     = {10.1145/2906142},
  url     = {https://www.plai.ifi.lmu.de/publications/csur16-obfuscation.pdf}
}

@inproceedings{du2024evaluating,
  title={Evaluating large language models in class-level code generation},
  author={Du, Xueying and Liu, Mingwei and Wang, Kaixin and Wang, Hanlin and Liu, Junwei and Chen, Yixuan and Feng, Jiayi and Sha, Chaofeng and Peng, Xin and Lou, Yiling},
  booktitle={Proceedings of the IEEE/ACM 46th International Conference on Software Engineering},
  pages={1--13},
  year={2024}
}

@inproceedings{pan2024lost,
  title={Lost in translation: A study of bugs introduced by large language models while translating code},
  author={Pan, Rangeet and Ibrahimzada, Ali Reza and Krishna, Rahul and Sankar, Divya and Wassi, Lambert Pouguem and Merler, Michele and Sobolev, Boris and Pavuluri, Raju and Sinha, Saurabh and Jabbarvand, Reyhaneh},
  booktitle={Proceedings of the IEEE/ACM 46th International Conference on Software Engineering},
  pages={1--13},
  year={2024}
}

@inproceedings{zhang2025unseen,
  title={Unseen horizons: Unveiling the real capability of llm code generation beyond the familiar},
  author={Zhang, Yuanliang and Xie, Yifan and Lit, Shanshan and Liu, Ke and Wang, Chong and Jia, Zhouyang and Huang, Xiangbing and Song, Jie and Luo, Chaopeng and Zheng, Zhizheng and others},
  booktitle={2025 IEEE/ACM 47th International Conference on Software Engineering (ICSE)},
  pages={604--615},
  year={2025},
  organization={IEEE}
}

@article{djire2026memorization,
  title={Learned or Memorized? Quantifying Memorization Advantage in Code LLMs},
  author={Euraste, Djir{\'e} Alb{\'e}rick and Kader, Kabor{\'e} Abdoul and Samhi, Jordan and Barr, Earl T and Klein, Jacques and Bissyand{\'e}, Tegawend{\'e} F},
  journal={arXiv preprint arXiv:2604.13997},
  year={2026}
}

@article{jiang2026cascade,
  title={Cascade: Llm-powered javascript deobfuscator at google},
  author={Jiang, Shan and Kovuri, Pranoy and Tao, David and Tan, Zhixun},
  journal={arXiv preprint arXiv:2507.17691},
  year={2025}
}

@article{pan2026readability,
  title={The hidden cost of readability: How code formatting silently consumes your llm budget},
  author={Pan, Dangfeng and Sun, Zhensu and Zhang, Cenyuan and Lo, David and Du, Xiaoning},
  journal={arXiv preprint arXiv:2508.13666},
  year={2025}
}

@article{yang2026membership,
  title={How Do Semantically Equivalent Code Transformations Impact Membership Inference on LLMs for Code?},
  author={Yang, Hua and Velasco, Alejandro and Le-Cong, Thanh and Haque, Md Nazmul and Xu, Bowen and Poshyvanyk, Denys},
  journal={arXiv preprint arXiv:2512.15468},
  year={2025}
}

@article{abdelsalam2026confused,
  title={Are Humans and LLMs Confused by the Same Code? An Empirical Study on Fixation-Related Potentials and LLM Perplexity},
  author={Abdelsalam, Youssef and Peitek, Norman and Maurer, Anna-Maria and Toneva, Mariya and Apel, Sven},
  year={2026}
}

@article{moller2012static,
  title={Static program analysis},
  author={M{\o}ller, Anders and Schwartzbach, Michael I},
  journal={Notes. Feb},
  year={2012},
  publisher={Addison-Wesley}
}

@article{roziere2023code,
  title={Code llama: Open foundation models for code},
  author={Roziere, Baptiste and Gehring, Jonas and Gloeckle, Fabian and Sootla, Sten and Gat, Itai and Tan, Xiaoqing Ellen and Adi, Yossi and Liu, Jingyu and Sauvestre, Romain and Remez, Tal and others},
  journal={arXiv preprint arXiv:2308.12950},
  year={2023}
}

@article{zhu2024deepseek,
  title={Deepseek-coder-v2: Breaking the barrier of closed-source models in code intelligence},
  author={Zhu, Qihao and Guo, Daya and Shao, Zhihong and Yang, Dejian and Wang, Peiyi and Xu, Runxin and Wu, Y and Li, Yukun and Gao, Huazuo and Ma, Shirong and others},
  journal={arXiv preprint arXiv:2406.11931},
  year={2024}
}

@article{guo2025deepseek,
  title={Deepseek-r1: Incentivizing reasoning capability in llms via reinforcement learning},
  author={Guo, Daya and Yang, Dejian and Zhang, Haowei and Song, Junxiao and Wang, Peiyi and Zhu, Qihao and Xu, Runxin and Zhang, Ruoyu and Ma, Shirong and Bi, Xiao and others},
  journal={arXiv preprint arXiv:2501.12948},
  year={2025}
}

@article{hui2024qwen2,
  title={Qwen2. 5-coder technical report},
  author={Hui, Binyuan and Yang, Jian and Cui, Zeyu and Yang, Jiaxi and Liu, Dayiheng and Zhang, Lei and Liu, Tianyu and Zhang, Jiajun and Yu, Bowen and Lu, Keming and others},
  journal={arXiv preprint arXiv:2409.12186},
  year={2024}
}

@misc{qwen2025qwen3coder30b,
  author       = {Qwen Team},
  title        = {Qwen3-Coder-30B-A3B-Instruct},
  year         = {2025},
  howpublished = {\url{https://huggingface.co/Qwen/Qwen3-Coder-30B-A3B-Instruct}},
  note         = {Hugging Face model card, accessed 2026-04-23}
}

@misc{openai2025gpt41,
  author       = {{OpenAI}},
  title        = {Introducing GPT-4.1 in the API},
  year         = {2025},
  howpublished = {\url{https://openai.com/index/gpt-4-1/}},
  note         = {Accessed 2026-04-23}
}

@article{cimato2005overcoming,
  title={Overcoming the obfuscation of Java programs by identifier renaming},
  author={Cimato, Stelvio and De Santis, Alfredo and Petrillo, U Ferraro},
  journal={Journal of systems and software},
  volume={78},
  number={1},
  pages={60--72},
  year={2005},
  publisher={Elsevier}
}

@inproceedings{venkatesh2024emergence,
  title={The emergence of large language models in static analysis: A first look through micro-benchmarks},
  author={Venkatesh, Ashwin Prasad Shivarpatna and Sabu, Samkutty and Mir, Amir M and Reis, Sofia and Bodden, Eric},
  booktitle={Proceedings of the 2024 IEEE/ACM First International Conference on AI Foundation Models and Software Engineering},
  pages={35--39},
  year={2024}
}

@article{codecipher,
  title={Codecipher: Learning to obfuscate source code against llms},
  author={Lin, Yalan and Wan, Chengcheng and Fang, Yixiong and Gu, Xiaodong},
  journal={arXiv preprint arXiv:2410.05797},
  year={2024}
}

@article{obfuscation,
  title={Code obfuscation literature survey},
  author={Balakrishnan, Arini and Schulze, Chloe},
  journal={CS701 Construction of compilers},
  volume={19},
  number={31},
  pages={10},
  year={2005}
}

@article{li2025systematic,
  title={A Systematic Study of Code Obfuscation Against LLM-based Vulnerability Detection},
  author={Li, Xiao and Li, Yue and Wu, Hao and Zhang, Yue and Zhang, Yechao and Xu, Fengyuan and Zhong, Sheng},
  journal={arXiv preprint arXiv:2512.16538},
  year={2025}
}

@article{nikiema2025code,
  title={The code barrier: What llms actually understand?},
  author={Nikiema, Serge Lionel and Samhi, Jordan and Kabor{\'e}, Abdoul Kader and Klein, Jacques and Bissyand{\'e}, Tegawend{\'e} F},
  journal={arXiv preprint arXiv:2504.10557},
  year={2025}
}

@inproceedings{beste2025exploring,
  title={Exploring the potential of llms for code deobfuscation},
  author={Beste, David and Menguy, Grgoire and Hajipour, Hossein and Fritz, Mario and Cin, Antonio Emanuele and Bardin, Sbastien and Holz, Thorsten and Eisenhofer, Thorsten and Sch{\"o}nherr, Lea},
  booktitle={International Conference on Detection of Intrusions and Malware, and Vulnerability Assessment},
  pages={267--286},
  year={2025},
  organization={Springer}
}

@article{hu2026can,
  title={Can LLMs Deobfuscate Binary Code? A Systematic Analysis of Large Language Models into Pseudocode Deobfuscation},
  author={Hu, Li and Shang, Xiuwei and Shi, Jieke and Cheng, Shaoyin and Zhang, Junqi and Li, Gangyang and Yang, Zhou and Zhang, Weiming and Lo, David},
  journal={arXiv preprint arXiv:2604.08083},
  year={2026}
}

@inproceedings{hort2025semantic,
  title={Semantic-preserving transformations as mutation operators: A study on their effectiveness in defect detection},
  author={Hort, Max and Vidziunas, Linas and Moonen, Leon},
  booktitle={2025 IEEE International Conference on Software Testing, Verification and Validation Workshops (ICSTW)},
  pages={337--346},
  year={2025},
  organization={IEEE}
}

@article{tkachenko2025deconstructing,
  title={Deconstructing Obfuscation: A four-dimensional framework for evaluating Large Language Models assembly code deobfuscation capabilities},
  author={Tkachenko, Anton and Suskevic, Dmitrij and Adolphi, Benjamin},
  journal={arXiv preprint arXiv:2505.19887},
  year={2025}
}

@article{le2025names,
  title={When Names Disappear: Revealing What LLMs Actually Understand About Code},
  author={Le, Cuong Chi and Pham, Minh VT and Van, Cuong Duc and Phan, Hoang N and Phan, Huy N and Nguyen, Tien N},
  journal={arXiv preprint arXiv:2510.03178},
  year={2025}
}

\end{document}